\documentstyle[psfig]{kluwer}

\runningtitle{ Fractal structures driven by self-gravity}
\runningauthor{ F. Combes}

\begin{opening}

\title{ FRACTAL STRUCTURES DRIVEN by SELF-GRAVITY:\\
Molecular clouds and the Universe}

\bigskip
\author{ Francoise \surname{Combes}}

\institute{Observatoire de Paris, DEMIRM\\
 61 Av. de l'Observatoire, F-75014 Paris, FRANCE}

\end{opening}

\begin{document}

\begin{abstract}
 In the interstellar medium, as well as in the Universe,
large density fluctuations are observed, that obey
power-law density distributions and correlation functions.
 These structures are hierarchical, chaotic, turbulent,
but are also self-organizing. The apparent disorder is not random noise, 
but can be described by a fractal, with a deterministic
fractal dimension.  We discuss the theories advanced to
describe these fractal structures, and in particular
a new theory of the self-gravity thermodynamics, that could explain 
their existence, and predict their
fractal dimension. The media obeying scaling laws
can be considered critical, as in second order phase transitions
for instance.
\end{abstract}


\bigskip
\section{Introduction: Fractal Concepts}

 Fractals have been introduced by Mandelbrot (1975) to define 
geometrical ensembles, or mathematical sets, that have a fractional
dimension. He pioneered the study of very irregular mathematical 
sets, where the methods of classical calculus cannot
be applied. Fractals are not smooth nor differentiable;
they are characterized by self-similarity. Their geometrical
structure has details at all scales, and the details are
representative of the whole. 

Fractals are very rich, since they give the best 
approximation for many natural phenomena, that 
cannot be represented by regular and smooth geometries
(natural systems are only fractal between two boundaries,
with upper and lower cut-offs, and not true mathematical 
fractals, of course).
There can exist a large part of randomness in fractals,
and the fractal dimension and exponents of
scaling laws are the tools to quantify the
hidden order in them.

There are many definitions of the dimension, trying
to quantify how much space a system fills.
We will use mainly the Hausdorff dimension $D$, 
based on the Hausdorff measure, that generalizes
the notion of length, area and volume. When the
length scale is magnified by $\lambda$, a regular and smooth
set's area and volume are respectively scaled by 
$\lambda^2$, $\lambda^3$, while the fractal set measure
is scaled by $\lambda^D$.
For the astrophysical systems that we describe below,
the mass contained within a scale $r$ is M$\propto r^D$,
with $D$ a fractional number between 1 and 3.

\bigskip
 \section{Interstellar Clouds}

The gaseous medium pervading the galaxy in between stars is highly
structured. This medium consists of clouds of hydrogen, either atomic
or molecular, according to its density or column density. A quick look at
the Milky Way reveals at once the very irregular and clumpy structure
of the interstellar medium (ISM), while the stellar distribution is much
smoother. Sizes of the clouds range from 100pc (the so-called Giant
Molecular Clouds or GMC) down to 20 AU the smallest structures observed
through VLBI in the vicinity of the Sun, in absorption in front of
quasars (e.g. Davis et al 1996). This corresponds to 6 orders of
magnitude in size, and about 10 in masses.

The geometrical appearance is very irregular, but can be caracterized
by sheets and filaments of great contrast. The aspect is self-similar,
which can be glanced from clouds in the Galaxy at very different distances
from the Sun. The dynamics of the ISM has always been mysterious, since
it was expected that clouds collapse in a few million years to form stars.
But molecular clouds, at all scales, are found to be relatively stable
over times long compared to the free-fall time of the cloud at the given scale.

\bigskip
\subsection{ Scaling Laws}

\begin{figure}[t]
\psfig{figure=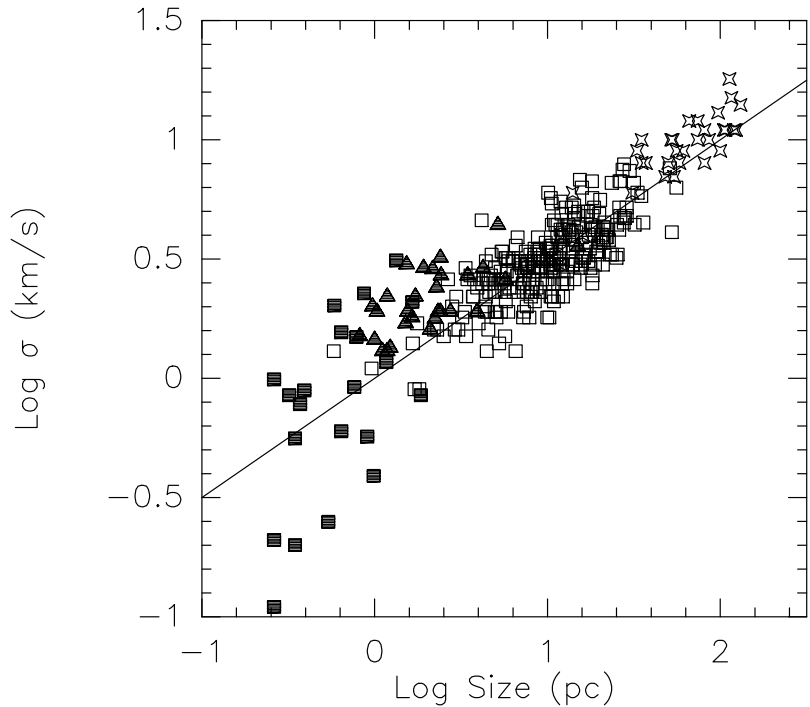,bbllx=65mm,bblly=5mm,bburx=185mm,bbury=145mm,width=12cm,angle=-90}
\caption{Size-linewidth relation taken from Solomon et al (1987) for the 
empty squares, from Dame et al (1986) for the empty stars, from Heithausen
(1996) for the filled triangles, and from Magnani et al (1985) for the
filled squares. The two latter samples concern high latitude molecular clouds.
}
\label{fig1}
\end{figure}

The various structures of interstellar clouds are not distributed at
random, but obey power-law relations between size, linewidth and mass
(cf Larson 1981). These power-laws demonstrate the self-similar nature,
and the scaling properties of the ISM: no peculiar scale exists (except
at the two boundaries of course, lower and upper cut-offs).
 The mass of the clouds is always a very uncertain quantity to obtain,
since there is no good universal tracer. The H$_2$ molecule does not
radiate in the cold conditions of the bulk of the ISM (10-15 K),
since it is symmetric, with no dipole moment. The first tracer is
the most abundant molecule CO (10$^{-4}$ with respect to H$_2$), but
it is most of the time optically thick, or photo-dissociated (cf below).
 More direct quantities to measure are the sizes $R$ and the line-widths 
or velocity dispersion $\sigma$, and the two are linked through a power-law 
relation:

$$
 \sigma \propto R^q
$$
with $q$ between 0.3 and 0.5 (e.g. Larson 1981, Scalo 1985, 
Solomon et al 1987, cf \ref{fig1}).
 Besides, molecular clouds appear to be virialised (at least within
the uncertainties of mass determination) over all scales, so that

$$
 \sigma^2 \propto M/R
$$
and the size-mass relation follows:

$$
M \propto R^D
$$
with $D$ the Hausdorff fractal dimension between 1.6 and 2.
It can be deduced also that the mean density over a given scale R
decreases as 1/R$^\alpha$, where $\alpha$ is between 1 and 1.4.

\bigskip
\subsection{  Hierarchical and Tree Interpretations}

The observations of molecular clouds reveal that the structure
is highly hierarchical, smaller clumps being embedded within the 
larger ones. Is this a completely exclusive relation, or are
there isolated clumps? This is difficult to disentangle, since
we have no real 3D picture of the ISM, the third dimension being
traced by the radial velocity, and the latter being turbulent and not
systematic. It has been possible, however, to build a tree structure
where each clump has a parent for instance for clouds in Taurus, a very 
nearby region (Houlahan \& Scalo 1992). 

This recursive structure might tell us about the formation mechanism,
as will be detailed later on. Basically, if self-gravity is the dominant 
force, density contrast will be build through Jeans instability, and
this is a recursive process, in a quasi isothermal medium. The cooling
is very efficient in the dense ISM, and it can be considered almost
isothermal. 

\bigskip
\subsection{  Fractal Dimensions}

There are various ways to quantify the observed self-similarity,
and to define fractal dimensions. One of them
is just to measure in 2D the surface versus the perimeter of a given
structure. This method has been used in 2D maps, like the IRAS continuum
flux, or the extinctions maps of the sky. In all cases, this
method converge towards the same fractal dimension $D_2$.
For a curve of fractal dimension $D_2$ in a plane, the perimeter P and area
A are related by
$$
 P \propto A^{D_2/2}
$$
Falgarone et al (1991) find a dimension $D_2$ = 1.36 for CO contours
both at very large (degrees) and very small scales (arcmin), and the same
is found for IRAS 100$\mu$ contours in many circumstances (e.g. Bazell 
\& Desert 1988). Comparable dimensions ($D_2$ between 1.3 and 1.5) are
found with any tracer, for instance HI clouds
(Vogelaar \& Wakker 1994). 

Note that the projection of a fractal of dimension $D$ is not mandatorily a 
fractal, but if it is one with dimension $D_p$
it is impossible a priori to deduce its fractal dimensions, 
except that 

$D_p = D $ if $D \leq 2$

$D_p = 2 $ if $D \geq 2$

(Falconer 1990).

\bigskip
\subsection{  Technical Biases }

 It is a very difficult task to trace quantitatively the fractal
structure of the ISM. Dense gas is molecular, and cold H$_2$ molecules
do not radiate (Combes \& Pfenniger 1997). 
The trace molecules such as CO are either 
optically thick, or not thermally excited (in low-density regions),
or photo-dissociated near ionizing stars. The large range of scales
is also a source of bias: the small scales are not resolved, and 
observed maps are smoothed out. This process, which confuses the
fractal with a diffuse medium of fractal dimension 3, can lead to
underestimates of the mass by factors more than 10 (e.g. simulations
in Pfenniger \& Combes 1994).

The mass spectrum of density fluctuations has often been studied as 
another way to characterise the ISM structure, and may be also to
predict the mass spectrum of stars that form within these clouds.
The differential mass spectrum dN/dm has been found to obey a power-law
$$
dN/dm \propto m^\gamma
$$
with $\gamma = -1.5\pm 0.2$, between masses of 1 to 10$^6$ M$_\odot$
(Casoli et al 1984, Solomon et al 1987, Brand \& Wouterloot 1995).
But this result has been obtained through molecular lines surveys.
 When extinction surveys are used, to determine a size spectrum
(that can be directly related to the mass spectrum), a much 
flatter power law is found, and this has been attributed
to occlusion (Scalo \& Lazarian 1996), i.e. due to blocking
of clouds by larger foreground clouds. Although this affects
much less spectral studies, there could also be occlusion
at a given velocity.

More recently, Heithausen et al (1998) have extended this relation
over 5 orders of magnitudes in masses, down to Jupiter masses,
and they found a steeper slope  $\gamma = -1.84$. Their mass-size
relation is M$\propto r^{2.31}$, also much steeper than previous
studies, but the estimation of masses at small scales is quite
uncertain (in particular the conversion factor between CO and H$_2$
mass could be much higher).

\bigskip
\subsection{  Turbulence}

Everybody agrees that the word characterising the best the ISM 
is "turbulence". In laboratory it is well known that a laminar flow
can turn turbulent when the Reynolds number is larger than a critical value,
i.e. 
$$
R_e = v l /\nu > R_c
$$
where $v$ is the velocity, $l$ a typical dimension, and $\nu$ the
kinematic viscosity. This means that the advection term v . $\nabla$ v
dominates the viscous term in the fluid equation. 
The turbulent state is characterised by
unpredictable fluctuations in density and pressure, and a cascade
of whirls. 
In the ISM, the viscosity can be estimated from the product of the
macroturbulent velocity (or dispersion) and the mean-free-path of cloud-cloud
collisions (since the molecular viscosity is negligible).
 But then the Reynolds number is huge ($\approx 10^9$), and the presence
of turbulence is not a surprise. 

This fact has encouraged many interpretations of the ISM structure
in terms of what we know from incompressible turbulence. 
In particular, the Larson relations have been found as a sign of the
Kolmogorov cascade (Kolmogorov 1941). In this picture, energy is dissipated
into heat only at the lower scales, while it is injected only at large
scale, and transferred all along the hierarchy of scales. Writing that 
the energy transfer rate $v^2/(r/v)$ is constant gives the relation
$$
v \propto r^{1/3}
$$
which is close to the observed scaling law, at least for the 
smallest cores (Myers 1983).
  The source of energy at large scale would then be the differential
galactic rotation and shear (Fleck 1981).  This idealized view has 
been debated (e.g. Scalo 1987): it is not obvious 
that the energy cascades down without
any dissipation in route (or injection), given the large-scale shocks, flows, 
winds, etc... observed in the ISM. Also, the interstellar medium is highly 
compressible, and its behaviour could be quite different from
ordinary liquids in laboratory.

Besides, many features of ordinary turbulence are present in the ISM.
For instance, Falgarone et al (1991) have pointed out that the
existence of non-gaussian wings in molecular line profiles might be
the signature of the intermittency of the velocity field in turbulent
flows. More precisely, the $^{13}$CO average velocity profiles
have often nearly exponential tails, as shown by the velocity derivatives
in experiments of incompressible turbulence (Miesch \& Scalo 1995).
Comparisons with simulations of compressible gas give similar results
(Falgarone et al 1994).
Also the curves obtained through 2D slicing of turbulent flows
have the same fractal properties as the 2D projected images of the ISM;
their fractal dimension D$_2$ obtained from the perimeter-area relation
is also 1.36 (Sreenivasan \& M\'eneveau 1986).

More essential, the ISM is governed by strong fluctuations in density
and velocity. It appears chaotic, since it obeys highly non-linear
hydrodynamic equations, and there is coupling of phenomena at all scales.
 This is also related to the sensibility to initial conditions that
defines a chaotic system. The chaos is not synonymous of random
disorder, there is a remarquale ordering, which is reflected in the
scaling laws. The self-similarity over several orders of magnitude
in scale and mass means also that the correlation functions behave
as power-laws, and that there is no finite correlation length.
 This characterizes critical media, experiencing a second order
phase transition for example. This analogy will be developped
further in this chapter.

\bigskip
\subsection{  Self-Gravity }

Although the ISM is a self-organizing, multi-scale medium, comparable
to what is found in laboratory turbulence, there are very special 
particularities that are not seen but in astrophysics. Self-gravity
is a dominant, while it has not to be considered in atmospheric clouds
for instance. It has been recognized by Larson (1981) and by many others
that at each scale the kinetic energy associated with the linewidths 
balances the gravitational energy: clouds are virialized approximately,
given their very irregular geometry.

This property of course has to stop at the largest scale, when the influence
of the galactic gravity, and associated shear, intervenes. This scale is
that of the Giant Molecular Clouds, of the order of 100pc in size, and
10$^6$ M$_\odot$ in mass. They are the largest self-gravitating structures
in the Galaxy. At the other extremity, the smallest scales are not well known.
 The observations through molecular lines, in nearby clouds with
millimetric interferometry, detect structures down to 0.01pc currently,
and even in high-latitude clouds distant by 100 pc, structures down
to 400 AU (2 10$^{-3}$ pc). Through VLBI, it is possible to go much
further. HI is detected and mapped in absorption in front of remote quasars
(Diamond et al 1989, Davis et al 1996), and contrasted structures of 20 AU
are ubiquitous. This small-scale structure is also traced by the 
interstellar scattering, and in particular extreme scattering events or ESE
(Fiedler et al 1987, 1994). The statistics on ESE is so large that
we know approximately the number of small clumps in the Galaxy: they are
about 1000 times more numerous than stars. If these objects are 
self-gravitating, at the end of the ISM hierarchy, their mass is of the order
of 10$^{-3}$ M$_\odot$, and they represent a significant mass component
of the Galaxy (Pfenniger \& Combes 1994). 

Self-gravity is widely accepted as dominant process in the ISM.
Gravitational collapse is accompanied by fragmentation in a system
with very efficient cooling, and this process can provide the
turbulent motions observed. The theory was first proposed by
Hoyle (1953) who showed that the isothermal collapse of a cloud
led to recursive fragmentation, since the Jeans length decreases
faster than the cloud radius. Rees (1976) has determined the size of
the smallest fragments, when they become opaque to their own 
radiation. They correspond roughly to the smallest scales observed
in the ISM (sizes of 10 AU, and masses of 10$^{-3}$ M$_\odot$,
see the physical parameters of the "clumpuscules" in 
Pfenniger \& Combes 1994).
Many other physical processes play a role in the turbulent ISM, as
for instance rotation and magnetic fields. But they cannot be identified as
the motor and the origin of the structure. Galactic rotation certainly
injects energy at the largest scales, but angular momentum cannot
cascade down the hierachy of clouds; indeed if the rotational velocity
is too high, the structure is unstable to clump formation (cf Toomre
criterium, 1964), and the non-axisymmetry evacuates angular momentum outside
the structure. Magnetic fields are certainly enhanced by the turbulent
motions, and could reach a certain degree of global equipartition
with gravitational and kinetic energies in tbe virialised clouds.
 But they cannot be alone at the origin of the hierarchical structure,
gravity has to trigger the collapse first. Besides, there is
no observational evidence of the gas collapse along the field lines,
polarisation measurements give contradictory results for the field
orientation with respect to the gas filaments. Therefore, although
rotation, turbulence, magnetic fields play an important role in the ISM,
they are more likely to be consequences of the formation of the structure.

\bigskip
\subsection{  Simulations }

A large number of hydrodynamical simulations have been run, in order
to reproduce the hierachical density structure of the interstellar
medium. However, these are not yet conclusive, since the dynamical
range available is still restricted, due to huge computational requirements.

It has been argued that self-similar statistics alone can generate the
observed structure of the ISM, in pressureless turbulent flows without 
self-gravity (Vazquez-Semadeni 1994); however, only three levels of
hierarchical nesting can be traced.

From the size-linewidth relation $ \sigma \propto R^{1/2}$, and the second
observed scaling law $ \rho \propto R^{-1}$, it can be deduced that

$$
\sigma \propto \rho^{-1/2}
$$
and therefore, if the turbulent pressure $P$ is defined as usual by

$$ dP /d\rho = \sigma^2 $$
it follows that
$$
P \propto log\rho
$$
which is the logatropic equation of state, or "logatrope".
This behaviour has been tested in simulations (e.g. Vazquez-Semadeni et al
1998), but the logatrope has not been found adequate to represent 
dynamical processes occuring in the ISM (either hydro, or magnetic 2D
simulations). The equation of state of the gas would be more similar
to a polytrope of index $\gamma \approx 2$. But the results could
depend whether the clouds are in approximate equilibrium or not
(cf McLaughlin \& Pudritz 1996).

Vazquez-Semadeni et al (1997) have searched for Larson relations in
the results of 2D self-gravitating hydro (and MHD) simulations
of turbulent ISM: they do not find clear relations, but instead a
large range of sizes at a given density, and a large range of column
densities; they suggest that the observational results could be
artefacts or selection effects (existence of a threshold in
column density for UV-shielding for example).

\bigskip
\section{  Galaxy Distributions}

It has long been recognized that galaxies are not distributed
homogeneously in the sky, but they follow a hierarchical 
structure: galaxies gather in groups, that are embedded in
clusters, then in superclusters, and so on (Shapley 1934, Abell 1958). 
Moreover, galaxies and clusters appear to obey scaling properties,
such as the power-law of the two point-correlation function:
$$
\xi(r) \propto r^{-\gamma}
$$
with the slope $\gamma$, the same for galaxies and clusters, of $\approx$ 1.7
(e.g. Peebles, 1980, 1993). 

\bigskip
\subsection{  Correlation Functions}

The correlation function is defined as
$$
\xi(r) = \frac{<n(r_i).n(r_i+r)>}{<n>^2} -1
$$
where $n(r)$ is the number density of galaxies, and $<...>$ is the volume
average (over $d^3r_i$). One can always define a 
correlation length $r_0$ by $\xi(r_0) = 1$.

This definition involves the average density $<n>$, which depends on
the scale for the galaxy distribution, since it is a fractal, at
least over a certain range of scales. There has been considerable
debate about this (see Davis 1997, Pietronero et al 1997). 
If everybody agrees that the universe is a fractal below 200 Mpc scales,
the question is not settled as of the scale
beyond which the universe is homogeneous. Pietronero et al
(1997) claim that this limiting scale has not yet been reached in 
the present catalogs, since large-scale structures are still found
at any scale.
On the contrary, Davis \& Peebles (1983) or Hamilton (1993) argue that
the galaxy-galaxy correlation length  $r_0$ is rather small. 
The most frequently reported value is $r_0 \approx 5 h^{-1}$ Mpc
(where $h = H_0$/100km s$^{-1}$Mpc$^{-1}$).

The problem is that the definition of $\xi(r)$ includes a normalisation by
the average density of the universe, which, if the homogeneity scale is not 
reached, depends on the size of the galaxy sample. \footnote{The notion of 
correlation length $\xi_0$ is usually different in physics,
where  $\xi_0$ characterizes the exponential decay of correlations $ (\sim 
e^{- r/ \xi_0} ) $. For power decaying correlations, it is said that the  
correlation length is infinite}.

This implies a correlation length that should increase 
with the distance limits of galaxy catalogs, as it
indeed does (Davis et al 1988).
The same problem occurs for the two-point correlation function of
galaxy clusters; the corresponding $\xi(r)$ has the same power law 
as galaxies, their  length  $r_0$ has been reported to be about 
$r_0 \approx 25 h^{-1}$ Mpc, and their correlation amplitude is therefore
about 15 times higher than that of galaxies
(Postman, Geller \& Huchra 1986, Postman, Huchra \& Geller 1992).
The latter is difficult to understand, unless there is a considerable
difference between galaxies belonging to clusters and field galaxies (or
morphological segregation). The other obvious explanation is that
the normalizing average density of the universe was then chosen lower.

Assuming that the average density is a constant, while homogeneity
is not yet reached, could perturb significantly the correlation 
function, and its slope, as shown by Coleman, Pietronero \& Sanders (1988) 
and Coleman \& Pietronero (1992).
 The function $\xi(r)$ has a power-law behaviour 
of slope $-\gamma$ for $r< r_0$, then it turns down to zero 
rather quickly at the statitistical limit of the sample. This rapid
fall leads to an over-estimate of the small-scale $\gamma$.
Pietronero (1987) introduces the conditional density
$$
\Gamma(r) = \frac{<n(r_i).n(r_i+r)>}{<n>} 
$$
which is the average density around an occupied point.
For a fractal medium, where the mass depends on the size as
$$
M(r) \propto r^D
$$ 
$D$ being the fractal (Haussdorf) dimension, the conditional
density behaves as
$$
\Gamma(r) \propto r^{D-3}
$$

\begin{figure}[t]
\psfig{figure=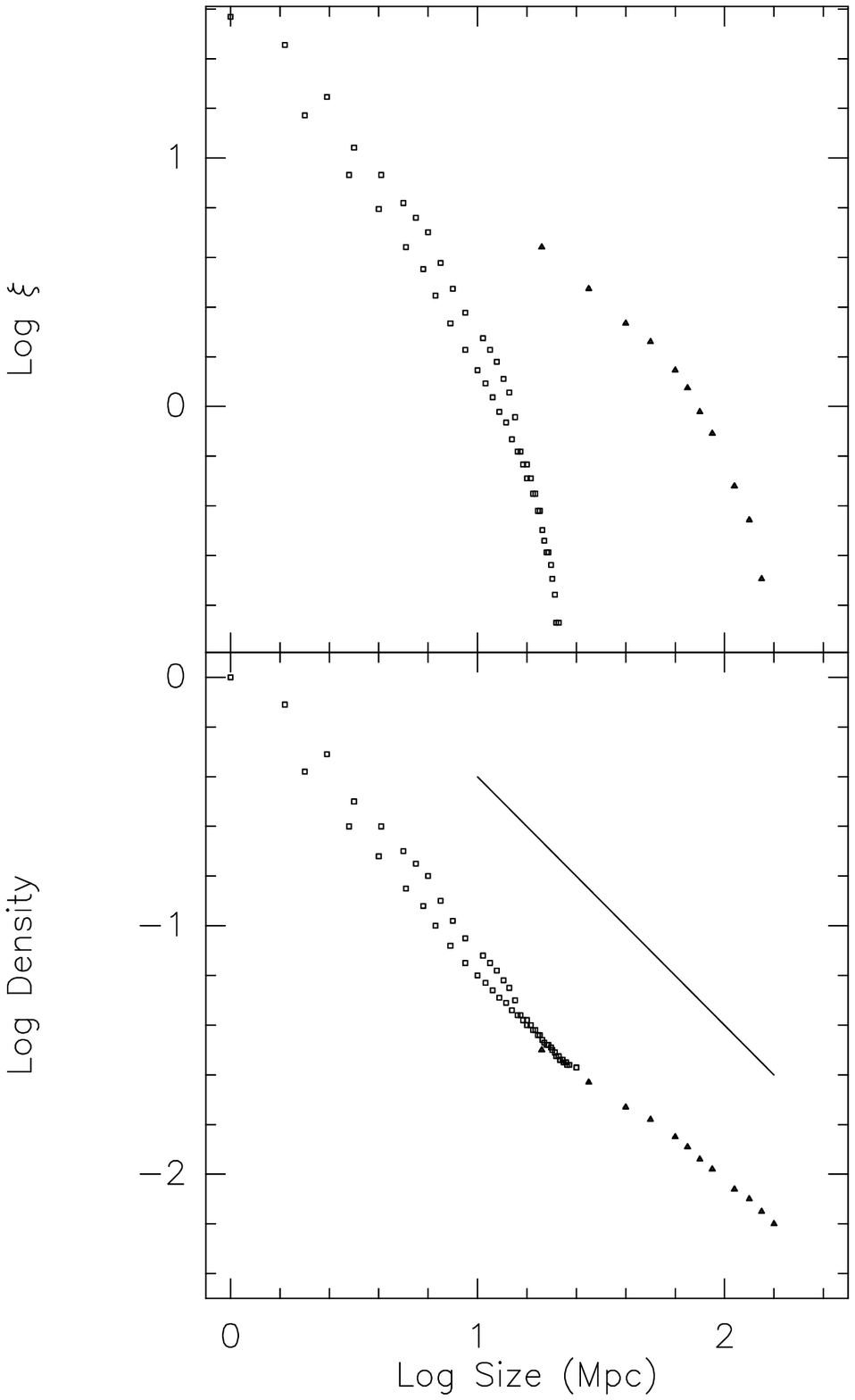,bbllx=1cm,bblly=2cm,bburx=15cm,bbury=24cm,width=10cm}
\caption{ {\it Bottom:} the average conditional density $\Gamma(r)$ for
several samples (Perseus-Pisces and CfA1 in open rectangles, and LEDA in
filled triangles, adapted from Sylos-Labini \& Pietronero 1996).
{\it Top:} $\xi(r)$ corresponding to the same surveys.
The indicative line has a slope $\gamma=1$ (corresponding to a fractal dimension
$D\sim2$). This shows that it is difficult to determine the slope on
the $\xi(r)$ function.}
\label{fig2}
\end{figure}

It is possible to retrieve the correlation function as
$$
\xi(r) = \frac{\Gamma(r)}{<n>} -1
$$
In the general use of
$\xi(r)$,  $<n>$ is taken for a constant, and we can see that
$$
D = 3 - \gamma \quad .
$$
If for very small scales,
both $\xi(r)$ and $\Gamma(r)$ have the same power-law behaviour, with the 
same slope $-\gamma$, then the slope appears to steepen for $\xi(r)$
when approaching the  length $r_0$. This explains why
with a correct statistical 
analysis (Di Nella et al 1996, Sylos Labini \& Amendola 1996, 
Sylos Labini et al 1996), the actual $\gamma \approx 1-1.5$ is smaller 
than that obtained using $\xi(r)$ (cf \ref{fig2}).
 This also explains why the amplitude of
$\xi(r)$ and $r_0$ increases with the sample size, and for clusters as well. 

\bigskip
\subsection{ Homogeneity Hypothesis and Cosmological Principle}

Isotropy and homogeneity are expected at very large scales from the
Cosmological Principle (e.g. Peebles 1993). However, this does not imply
local or mid-scale homogeneity (e.g. Mandelbrot 1982, Sylos Labini 1994):
a fractal structure can be locally isotropic, but inhomogeneous.
The main observational evidence in favor of the Cosmological Principle
is the remarkable isotropy of the
cosmic background radiation (e.g. Smoot et al 1992), that provides information
about the Universe at the matter/radiation decoupling. There must therefore
exist a transition between the small-scale fractality to large-scale
homogeneity. This transition is certainly smooth, and might correspond to the
transition from linear perturbations to the non-linear gravitational collapse 
of structures. The present catalogs do not yet see the transition since
they do not look up sufficiently back in time. It can be noticed that
some recent surveys begin to see a different power-law
behavior at large scales ($\lambda \approx 200-400  h^{-1}$ Mpc, e.g. 
Lin et al 1996).

\bigskip
\subsection{The Problem and Methods}

It is generally recognized that the galaxy structures in the Universe
have developped by gravitational collapse from primordial fluctuations.
Once unstable, density fluctuations do not grow as fast as we are used
to for Jeans instability (exponential), since they are slowed down
by expansion. The rate of growth is instead a power-law. Let us call 
the density contrast $\delta$:
$$ \delta(\vec x) = (\rho(\vec x) -<\rho>)/<\rho> $$
where $<\rho>$ is the mean density of the Universe, assumed
homogeneous at very large scale.
If $\vec r$ is the physical coordinate, the comoving coordinate 
$\vec x$ is defined by:
$$ \vec r = a(t) \vec x $$
where a(t) is the scale factor, accounting for the Hubble expansion
(normalised to a(t$_0$) = 1 at the present time). Since the Hubble constant
verifies $H(t) = \dot{a}/a$, the peculiar velocity is defined by
$$ \vec v = \dot{\vec r} - H \vec r = a \dot{\vec x}$$
In comoving coordinates, the Poisson equation becomes:
$$ \nabla_x^2\Phi = 4\pi G a^2 (\rho -<\rho>)$$
 It can be shown easily that in a flat universe, the density contrast
in the linear regime grows as the scale factor $a(t) = (t/t_0)^{2/3}$
(this is also approximately true for any universe in the early times).
 An exact solution for the non-linear collapses exist only in 
very special conditions, such as the spherical collapse. For the well-known
top-hat perturbation, an overdensity reaches the singularity in a finite
collapse time $t_c$, when its corresponding 
linear density contrast would have reached the value
$$\delta_c \approx 1.69$$
The evolution of its radius follows the Friedman solution for a density
above critical; it reaches a maximum radius, before virializing to
a radius equal to half that one. Its final density contrast is 178
(both figures, although indicative, are widely used in the domain).

 The level of fluctuations was very small (about 10$^{-5}$ at the scale 
of COBE resolution, i.e. 7$^\circ$) at the last scattering surface,
just before matter recombination, about 10$^5$ yr after the Big-Bang.
At first, the development of the structures is easy to compute, since
they are in the linear regime, and interactions between scales can
be neglected (cf Peebles 1980). Then, in the non-linear regime, 
no analytical solution exists, and one should resort to approximations,
or full N-body simulations.

\subsubsection{ Numerical simulations}

N-body computations
have been widely used, to gravitationnally
follow the non-linear evolution of the fluctuations.
From a comparison of the results with today galaxy distribution,
one can hope to trace back 
the initial mass spectrum of fluctuations, and to test 
postulated cosmologies such as CDM and related variants (cf Ostriker 1993). 
However, the evolution depends on many free parameters,
the gas physics, the star formation feedback, the amount
and physics of dark matter. This approach 
has not yet yielded definite results, also because numerical limitations
(restricted dynamical range due to the softening and limited volume) have
often masked the expected self-similar behavior (Colombi et al 1996).

\subsubsection{ Zel'dovich approximation }

The first approximation
has been pioneered by Zel'dovich (1970): it consists
to assume that the pressure forces are negligible in the first
collapsing structures, since they are much bigger than the Jeans length.
 This is particularly adapted to the top-down scenario of adiabatic
fluctuations, where photons and baryons both fluctuate, such that 
entropy is conserved.  In that case, the matter-photon coupling will 
prevent the development of large amplitudes fluctuations at small-scales.
 Large-scales are then the first to collapse, and they fragment in
smaller structures. In this approximation, the particles maintain their
peculiar velocities in comoving space, and collapse in 1D filaments,
or 2D sheets or pancakes (cf formation of caustics). 
After collapse, the approximation fails
since particles diffuse away; an artificial viscosity has then been 
introduced, so that particles cannot cross each other, and the pancakes
remain coherent, it is the "adhesion model" (Shandarin \& Zel'dovich 1989).
 The velocity field is governed by a Burgers equation, for which 
analytical solutions are known (Vergassola et al 1994).

Related to this approach is the Lagrangian approximation, which 
in fact pushes the Zel'dovich approximation to further orders.
 Since this approach fails as soon as orbits are crossing (i.e.
multi-streaming) it is assumed that this multi-streaming at small
scales has negligible consequences at large-scales (Lachi\`eze-Rey 1993).

The success of this approximation has been boosted by the fact that 
the self-gravitating Universe appears filamentary. The reason for this
is probably that many collapsing structures are larger than the Jeans mass;
also in realistic scenarios, the
collapse is not spherical, but does form sheets and filaments (Larson
1985). The latter have the advantage that in 1D, pressure forces have
always the same dependence with radius than the gravity forces, and
therefore cannot halt the collapse, whatever the cooling (Uehara et
al.\ 1996; Inutsuka \& Miyama 1997).

\subsubsection{ BBGKY hierarchy }

A second approach, which 
should work essentially in the linear  (or weakly non-linear) regime, 
is to solve the BBGKY hierarchy of equations, a method which
has been successfully used in plasma physics. However, the
hierachy is here infinite, and approximations should be made for
closure (Davis \& Peebles 1977; Balian \& Schaeffer 1989).
The main assumption is that the $N$-points 
correlation functions are
scale-invariant and behave as power-laws like is observed for the few-body
correlation functions. Crucial to this approach is the determination of
the void probability, which is a series expansion of the 
$N$-points correlation 
functions (White 1979). The hierachical solutions found in this frame
agree well with the simulations, and with the fractal structure of the universe
at small-scales (Balian \& Schaeffer 1988). 

\subsubsection{ Thermodynamical approach }

A third approach
is the thermodynamics of gravitating systems, developped
by Saslaw \& Hamilton (1984), which assumes quasi thermodynamic
equilibrium. The latter is justified at the small-scales of non-linear
clustering, since the expansion time-scale is slow with respect to local
relaxation times. Indeed the main effect of expansion is to subtract 
the mean gravitational field, which is negligible for structures of
mean densities several orders of magnitude above average.
  The predictions of the thermodynamical theory have been successfully
compared with N-body simulations (Itoh et al 1993), but a special physical
parameter (the ratio of gravitational correlation energy to thermal energy) 
had to be adjusted for a better fit (Bouchet et al 1991, Sheth \& Saslaw 1996,
Saslaw \& Fang 1996). 

\bigskip
\subsection{ Mass Function, Press-Schechter Formalism }

In 1974, Press and Schechter developped a formalism to predict the
mass function of galaxies and structures in the Universe, 
assuming that gravitational collapse in an expanding universe
non-linearly evolves through a self-similar spectrum (PS74). 
The resulting mass function is independent of the initial
spectrum of condensations assumed. The non-linear N-body 
interactions randomize the initial positions and generate
perturbations to all other scales: the growth of large scale
condensations from non-linear clumping of smaller ones
occuring much faster than their linear growth,
a self-similar universal spectrum is established,
only from seeds corresponding to the Jeans mass at
recombination ($\approx$ 10$^7$ M$_\odot$).
 
This led to an excellent fit of the galaxy 
luminosity function proposed by Schechter (1976):
$$
\Phi(L) \propto L^{-\alpha} exp(-L/L_*)
$$
where $L_*$ is a characteristic galaxy luminosity, and $\alpha\sim 1$.

In order to derive the mass function, i.e. the number density of structures 
of mass between M and M+dM, $n(M)$, PS74 first assume that a structure
collapses as soon as the extrapolated linear density contrast $\delta_c$ is of
the order of 1. Since in the spherical top-hat case (see section 3.3)
the collapse occurs in a time-scale corresponding to a linear $\delta_c$ 
= 1.69, this indicative value of $\delta_c$ is taken. The
fraction of collapsed mass, at any resolution 
smaller than $R$ is:

$$F(<R) = \int_{\delta_c} ^\infty P(\delta, R) d\delta$$

For gaussian random phase fluctuations, the probability $ P(\delta, R)$ is
$$ P(\delta, R) = (2\pi)^{-1/2}/\delta_* \, exp(-\delta^2/(2 \delta_*^2))$$

where $\delta_*$ is the standard deviation, obtained at scale $R$
$$\delta_* = (<M^2> - <M>^2)^{1/2}/M$$

Then the fraction $F(<R)$ becomes:
$$ F(<R) = 1/2 \, {\rm erfc}\, (2^{-1/2} \delta_c/ \delta_*)$$
where $erfc$ is the complementary error function.

To relate this to the mass associate to the scale R, we use
$ M = 4\pi <\rho> R^3/3$, justified for the top-hat smoothing
function (to obtain the average $\rho$).
The function $F(<R)$ gives the probability for a scale $R$ to be
bound, but it could also be part of another larger bound scale,
so the fraction of independent masses,
with mass between $M$ and $M+dM$,
 that have collapsed is
the derivative of it with respect to mass $M$:
$$ dF/dM = 1/2 (2\pi)^{-1/2} \delta_c /\delta_* M^{-1}
\, exp(-\frac{1}{2} \frac{\delta_c^2}{\delta_*^2} )$$

The mass density due to condensations of mass $M$ is then obtained by
multiplying by the total number density at the same epoch $\rho/M$.
 However, PS74 realised that the integral of the fraction F was 1/2
instead of 1, i.e. that the formula accounted for only one half of
the mass in the universe. They multiplied by a factor 2 to include the
underdense regions that would accrete on the collapsed objects.
This is a typical problem of the linear theory, where only half of the mass
is in overdense regions. Also called the cloud-in-cloud problem,
this factor 2 puzzle has been better explained later 
(Cole 1989, Bond et al 1991, White \& Kauffman 1994).

This formalism has been recently widely used, since it gives excellent
agreement with N-body simulations, and provides a way to compute
easily the merging histories of galaxies (Lacey \& Cole 1993, 1994).

If we make the particular assumption that the power spectrum of the 
fluctuations $P(k)$ (i.e. the square of the 
module of $\delta_k$ the Fourier transform of $\delta$)
is $P(k) \propto k^n$, then $\delta_* \propto M^{-(n+3)/6}$, and 
the PS formula becomes
$$ n(M) dM = <\rho> /M_*^2 /sqrt(2\pi) (n+3)/6 (M/M_*)^{(n+3)/6-2} $$
$$exp (-1/2 (M/M_*)^{(n+3)/3}) dM$$
where $M_*$ is the mass corresponding to the scale at which
the amplitude $\delta_*$ reaches $\delta_c$.

\bigskip
\subsection{ Multifractals}

 The idea of fractals as a good representation of the self-similar
clustering hierachy of galaxies has been proposed very early
(de Vaucouleurs 1960, 1970; Mandelbrot 1975). Since then, many authors have
shown that a fractal distribution indeed reproduces quite well the 
aspect of galaxy catalogs, for example by simulating a fractal and observing
it, as with a telescope (Scott, Shane \& Swanson, 1954; Soneira \& Peebles 
1978). When going into details, however, the distribution of galaxies might
appear more complex: first of all, catalogs are weighted by the
luminosity distribution, and not the mass distribution, and these
are not equivalent. For instance, there is a luminosity segregation
in clusters of galaxies, and slightly different fractal dimensions
can be obtained with different galaxy-types (elliptical/spirals/dwarfs). 
Also determination of masses of structures through different ways
(X-rays, gravitational lensing, etc..) suggest that the dark matter
is not exactly distributed as light, and there could be a significant
bias. All this has led to the introduction of multifractality
to represent the Universe (e.g. Sylos-Labini \& Pietronero 1996).
In a multifractal system, local scaling properties slightly
evolve, and can be defined by a continuous distribution of exponents.
This is a mere generalisation of a simple fractal, that links the space
and mass distributions. 
Mutlifractality may also better account for the transition to homogeneity,
with a fractal dimension varying with scale
(Balian \& Schaeffer 1989, Castagnoli \& Provenzale 1991, 
Martinez et al 1993, Dubrulle \& Lachi\`eze-Rey 1994).

\bigskip
\section{Bases for a Statistical Field theory}

Both for the ISM and for galaxy distributions in the Universe,
self-similar structures are observed over large ranges in scales.
 Scaling laws are observed, which translate by an average density
decreasing with scale as a power-law, of slope $-\gamma$ between -1.5 and -1,
corresponding to a fractal dimension $D = 3-\gamma$ between 1.5 and 2.

We propose in the following that self-gravity is a dominant factor
in the two media, and try to establish a statistitical theory
in the hope to explain the fractal structure. The theory should 
not only account for the existence of the structure, but also
be able to predict its fractal dimension and others critical exponents
(de Vega, Sanchez \& Combes, 1996a,b, 1998). 

Since gravity is scale-independent, there are opportunities for 
a mechanism to propagate over scales in a self-similar fashion.
For the ISM, in a quasi isothermal regime, a fractal structure
could be build through recursive Jeans instability and fragmentation. 
 This recursive fragmentation proceeds until the density is high enough to
reach the adiabatic regime. Self-gravity could be the principal 
origin of the fractal, with generated turbulent motions in virial
equilibrium at each scale. For galaxy formation, the smallest 
structures collapse first, and these influence the largest scale
in a non-linear manner. It is obvious that in both cases, the system 
does not tend to a stationary point, but develops fluctuations
at all scales, and these must be studied statistically.

In order to study its thermodynamics properties,
 we develop the grand partition function of the ensemble of
self-gravitating particles (section \ref{stat}). In transforming
the partition function through a functional integral
(section \ref{functional}), it can be shown that 
the system is exactly equivalent to a scalar field theory
(section \ref{scalar}). The theory
does not diverge, since the system is considered only between two
scale limits: the short-scale and large-scale cut-offs.
Through a perturbative approach it can be demonstrated that the system 
has a critical behaviour, for any parameter
(effective temperature and density). That is, we can consider the
self-gravitating gaseous medium as correlated at any scale, as for
the critical points phenomena in phase transitions 
(as was first suggested by  Totsuji \& Kihara 1969).

Since scaling behaviours are the best studied through renormalization
group theory, we use these methods to derive by analogy the critical
exponents of the system 
(section \ref{renorm}). The aim, that will be developped more fully 
in the following, is to relate the critical exponents of 
 well known universal critical phenomena, to the Haussdorf 
dimension of the astrophysics fractals.

\bigskip
\subsection{  Hamiltonian of the Self-Gravitating  Ensemble of N-bodies}
\label{stat}
 
  Let us consider a gas of particles submitted only to their self-gravity,
in thermal equilibrium at temperature T  $(kT = \beta^{-1}) $.
In the interstellar medium, quasi isothermality is justified,
due to the very efficient cooling. For unperturbed gas in 
the outer parts of galaxies, gas is in equilibrium with the cosmic background
radiation at $T \approx 3K$ (Pfenniger et al 1994, Pfenniger \& Combes 1994).
For a system of collapsing structures in the universe, this
can be a valid approximation, as soon as the gradient of
temperature is small over a given scale.

This isothermal character is essential for the description of the
gravitational systems as critical systems, as will be shown
later, so that the canonical ensemble appears the best adapted system.
We do not consider isolated gravitational systems, for which the microcanonical
system is generally used (e.g. Horwitz \& Katz 1978a,b; Padmanabhan 1990).
However, for easy mathematical development, we develop the partition function
in the grand canonical ensemble, allowing for a variable
number of particles $N$ (for application to real systems,
constant masses will later be considered).
The grand partition function ${\cal Z}$ and the Hamiltonian $H_N$ are

$$
{\cal Z} = \sum_{N=0}^{\infty}\; {{z^N}\over{N!}}\; \int\ldots \int
\prod_{l=1}^N\;{{d^3p_l\, d^3q_l}\over{h^3}}\; e^{- \beta H_N}
$$

$$
H_N = \sum_{l=1}^N\;{{p_l^2}\over{2m}} - G \, m^2 \sum_{1\leq l < j\leq N}
{1 \over { |{\vec q}_l - {\vec q}_j|}}
$$
where $z$ is the fugacity = $exp(-\beta \mu_c)$ in terms of the
gravito-chemical potential $\mu_c$.

This can be transformed, with continuous density
$\rho({\vec r})= \sum_{j=1}^N\; \delta({\vec r}- {\vec q}_j)\; $

$$
 \frac12 \, \beta G \, m^2 \sum_{1\leq l \neq j\leq N}
{1 \over { |{\vec q}_l - {\vec q}_j|}} =  \frac12\,  \beta \, G \, m^2
\int_{ | {\vec x} - {\vec y}|> a}\;
{{d^3x\, d^3y}\over { | {\vec x} - {\vec y}|}}\; \rho({\vec x})
\rho({\vec y}) \; 
$$

The cutoff $ a $ is here introduced naturally, it corresponds
to the size of the smallest fragments, or clumpuscules
(of the order of $\sim 10$ AU). In fact, we consider that the
particles of the system interact with the Newton
law of gravity ($1/r$) only within the size range of the fractal,
where self-gravity is predominant. At small scale, other forces
enter into account, and we can adopt a model of hard spheres
to schematize them. Also at large scales, beyond
the upper cutoff, different forces must be introduced.
The phenomenological potential thus considered does not possess
any singularity.

\bigskip
\subsection{  Grand Partition Function as a Functional}
\label{functional}

Using the potential in $1/r$, and its inverse operator 
$-\frac{1}{4\pi}\nabla^2$
(but see also a similar derivation, with $[1 - \theta (a-r) ]/r$ 
and its corresponding inverse operator,
for the phenomenological potential with cutoff, in
de Vega et al 1996b),
the exponent of the
potential energy can be represented 
as a functional integral (Stratonovich 1958, Hubbard 1959)

$$
e^{  \frac12\, \beta G \, m^2
\int \;
{{d^3x\, d^3y}\over { | {\vec x} - {\vec y}|}}\; \rho({\vec x})
\rho({\vec y})} = \int\int\; {\cal D}\xi \; e^{ -\frac12\int d^3x \; (\nabla
\xi)^2 \; + \; 2 m \sqrt{\pi G\beta}\; \int d^3x \; \xi({\vec x})\;
\rho({\vec x}) } 
$$

\begin{eqnarray}
{\cal Z} &=& \sum_{N=0}^{\infty} {1 \over{N!}}
\left [z\left({2\pi m \over{h^2 \beta}}\right)^{3/2}\right]^N\
 \int\int {\cal D}\xi  e^{ -\frac12\int d^3x  (\nabla \xi)^2}
\int\ldots \int
\prod_{l=1}^N d^3q_l e^{ 2 m \sqrt{\pi G\beta} \sum_{l=1}^N
\xi({\vec q}_l)} \cr \cr
 &=& \int\int {\cal D}\xi  e^{ -\frac12\int d^3x (\nabla \xi)^2}
 \sum_{N=0}^{\infty}{1 \over{N!}}
\left [ z\left({2\pi m \over{h^2 \beta}}\right)^{3/2}\right]^N
\left[ \int d^3q  e^{ 2 m \sqrt{\pi G\beta} \xi({\vec q})}
\right]^N \cr \cr
 &=& \int\int\;  {\cal D}\xi \; e^{ -\int d^3x \left[ \frac12(\nabla \xi)^2\;
- z \left({m \over{2\pi \beta}}\right)^{3/2}\; e^{ 2 m \sqrt{\pi
G\beta}\;\xi({\vec x})}\right]}\nonumber
\end{eqnarray}

With the following change of variables:
$$
\phi({\vec x}) \equiv  2 m \sqrt{\pi G\beta}\;\xi({\vec x}) 
$$

$$
{\cal Z} =  \int\int\;  {\cal D}\phi\;  e^{ -{1\over{T_{eff}}}\;
\int d^3x \left[ \frac12(\nabla\phi)^2 \; - \mu^2 \; e^{\phi({\vec
x})}\right]}
$$
where
$$
\mu^2 = {{\pi^{5/2}}\over {h^3}}\; z\; G \, (2m)^{7/2} \, \sqrt{kT} \; 
\quad , \quad T_{eff} = 4\pi \; {{G\; m^2}\over {kT}} \quad 
$$
(note that the "equivalent" temperature in the field theory
is in fact inversely proportional to the physical temperature).
It can be shown that the parameter $\mu$ is equal to the inverse
of the Jeans length, itself of the order of the cutoff $a$.

\bigskip
\subsection{ Introduction of a Local Scalar Field, with
Exponential Self-interaction}
\label{scalar}

The main point of the previous derivation is that the
partition function for the gas of particles in gravitational
interaction has been transformed into the partition function for a single
scalar field $\phi({\vec x})$  with  local action

$$
S[\phi(.)] \equiv  {1\over{T_{eff}}}\;
\int d^3x \left[ \frac12(\nabla\phi)^2 \; - \mu^2 \; e^{\phi({\vec
x})}\right] 
$$

Apparently, the term $ - \mu^2 \; e^{\phi({\vec x})} $
makes the  $\phi$-field energy density diverge, which is related
to the attractive
character of the gravitational force. But the physical short-distance
cutoff $a$ eliminates the zero distance singularity.
It is then possible to compute the statistical average  value of the density
$\rho({\vec r})$ 

$$
<\rho({\vec r})> = {\cal Z}^{-1} \sum_{N=0}^{\infty} {1 \over{N!}}
\left [ z\left({m \over{2\pi \beta}}\right)^{3/2}\right]^N
\int\int \prod_{l=1}^N d^3q_l\rho({\vec r}) 
e^{ \frac12 \beta G  m^2 \sum_{1\leq l \neq j\leq N}
{1 \over { |{\vec q}_l - {\vec q}_j|}} }
$$

Or in the $\phi$-field language, the particle density 
expresses as
$$
  <\rho({\vec r})> =  -{1 \over {T_{eff}}}\;<\nabla^2 \phi({\vec r})>=
{{\mu^2}\over{T_{eff}}} \; <e^{\phi({\vec r})}>  
$$
where $ <\ldots > $ means functional average over   $ \phi(.) $
with statistical weight $  e^{S[\phi(.)]} $. Density correlators are

$$
C({\vec r_1},{\vec r_2}) \equiv
<\rho({\vec r_1})\rho({\vec r_2}) > -<\rho({\vec r_1})><\rho({\vec r_2}) > 
$$
$$
C({\vec r_1},{\vec r_2}) =
  {{\mu^4}\over{T_{eff}}^2} \; \left[  
<e^{\phi({\vec r_1})} \; e^{\phi({\vec r_2})}> -
<e^{\phi({\vec r_1})}> \; <e^{\phi({\vec r_2})}> \right] 
$$

\bigskip
\subsection {Stationary Points and Hydrostatic Solutions}

The equation for stationary points:
$$
\nabla^2\phi = -\mu^2\,  e^{\phi({\vec x})} \; 
$$
can be expressed
in terms of the gravitational potential $U({\vec x})$ 
$$
\nabla^2U({\vec r}) = 4 \pi G \, z \,  m
\left({{2\pi mkT}\over{h^2}}\right)^{3/2} \,  e^{ - \frac{m}{kT}\,U({\vec r})} 
$$
This corresponds to the Poisson equation for 
an ideal gas in hydrostatic equilibrium:
$$
\nabla P({\vec r}) = - m \, \rho({\vec r}) \; \nabla U({\vec r})
$$
where $ P({\vec r}) $ stands for the pressure. Combined with the
equation of state for the ideal gas
$$
P = kT \rho 
$$
this yields for the particle density
$$
 \rho({\vec r}) =  \rho_0 \; e^{ - \frac{m}{kT}\,U({\vec r})} 
$$
where $ \rho_0 $ is a constant. 

The  $\phi$-field has special properties under scale transformations
$$
{\vec x} \to {\vec x}_{\lambda} \equiv \lambda{\vec x} 
$$
where $\lambda$ is an arbitrary real number. For any solution  $
\phi({\vec x}) $ of the stationary point equations,
there is a family of dilated solutions of the same equation
$$
\phi_{\lambda}({\vec x}) \equiv \phi(\lambda{\vec x}) +\log\lambda^2
$$
In addition, $ S[\phi_{\lambda}(.)] = \lambda^{-1} \; S[\phi(.)] $. 
A rotationally invariant stationary point is given by
$$
\phi^c(r) = \log{{2}\over { \mu^2 r^2}} 
$$
This singular solution, where can be recognized the isothermal
sphere, is invariant under the scale
transformations, i.e.
$$
\phi^c_{\lambda}(r) =\phi^c(r)
$$
The only constant stationary solution is
the singular $ \phi_0 = -\infty $.

With a perturbative method,
starting from the stationary solution $ \phi_0 = -\infty $,
it can be obtained for large distances (de Vega et al 1996b)
$$
 C({\vec r_1},{\vec r_2}) \buildrel{  | {\vec  r_1} - {\vec  r_2}|\to
\infty}\over =  {{ \mu^4 }\over {2\, C_D^2 \; 
 | {\vec  r_1} - {\vec  r_2}|^{2}}} + O\left( \; | {\vec
r_1} - {\vec r_2}|^{-3}\right)
$$
showing that the $\phi$-field theory  scales, and 
behaves  critically for a continuum set of  values of $\mu$ and
$ T_{eff} $.

\bigskip
\section{ Renormalization Group Methods}
\label{renorm}

The renormalization methods are very powerful to deal with
self-similar systems obeying scaling laws, like critical
phenomena. In the latter case, examplified by second 
order phase transitions, there exist critical divergences,
where physical quantities become singular as power-laws
of parameters called critical exponents.
These critical systems reveal a collective behaviour,
organized from microscopic degrees of freedom, through giant
fluctuations and statistical correlations.
Hierarchical structures are built up, coupling all scales
together, replacing an homogeneous system
in a scale-invariant system. It can be shown that 
local forces are not important to describe the collective behaviour,
which is only due to the statistical coupling of local
interactions. Therefore, critical exponents depend only
on the statistical distribution of microscopic
configurations, i.e. on the dimensionalities or
symmetries of the system. There exist wide universality
classes, that allow to draw quantitative predictions
on the system from only a qualitative knowledge of
its properties (e.g. Parisi, 1988; Zinn-Justin 1989; Binney et al 1992).

\bigskip
\subsection{  Critical Phenomena}

Critical phenomena occur at second order phase transitions,
i.e. continuous transitions without latent heat. The paradigm of
these systems is the transition at the Curie point (T=T$_c$=
1043K) from paramagnetic iron, where the magnetic moment 
is proportional to the applied field m=$\mu$ B, to ferromagnetic
state, where there exists a permanent magnetic moment m$_0$
even in zero field. Another well known example is the
critical point of water, at which the transition from the liquid
to gas becomes continuous (at T$_c$ = 647 K, $\rho_c$ = 0.323 g cm$^{-3}$).

Although the permanent magnet tends to zero continuously at T$_c$,
there are divergences: for instance the heat capacity C
behaves as C$ \propto | T - T_c |^{-\alpha}$, with $\alpha > 0$.
Also the magnetic susceptibility
$$ \chi_T = \partial{m}/\partial{B}_T \propto  | T - T_c |^{-\gamma}$$
(or for the case of H$_2$O, the compressibility $\kappa_T \propto
 | T - T_c |^{-\gamma}$).

At the critical point, it is easy to understand that
the compressibility which tends to infinity generates
large density fluctuations, and therefore light is
strongly diffused by the varying optical index:
this is the critical opalescence. The extraordinary
fact is that microscopic forces can give rise
to large-scale fluctuations, as if the medium was
organized at all scales. 

\subsubsection{ Order parameter and correlation function}

The order parameter is defined by the characteristic 
physical quantity which experiences large fluctuations
at criticality. For the H$_2$O case, it is a scalar
field $$ \Phi(x) = \rho(x) -\rho_{gas}(x)$$
for a spin system, it is a vector field
($\Sigma s_i$), according to the dimensionality
adopted. The critical point is particularly
well characterized through the correlation functions.
 The two-point correlation function is 
$$ G^{(2)} (r) = < \Phi(0). \Phi(r) >$$
where brackets mean statistical (thermal) average,
over all configurations, and the connected
correlation function is 
$$ G^{(2)}_c (r) = < \Phi(0). \Phi(r) > - |<\Phi>|^2 $$
which is independent of the mean value.
At critical point, for large distances $r$
$$ G^{(2)}_c (r) \propto r^{2-d-\eta} $$
where $d$ id the space dimension, and $\eta$
a critical exponent. Far from the critical point,
$$ G^{(2)}_c (r) \propto \, exp(-r/\xi) $$
where $\xi \propto  | T - T_c |^{-\nu}$ is the
correlation length, and $\nu$ another critical
exponent.

\subsubsection{  Universality, Dimensionality, Symmetry}

Experiments have shown that the critical exponents
for a wide variety of systems are the same, and more
precisely they belong to universality classes, depending
only on the dimensionality $d$ of space and $D$ of
the order parameter (for instance if the field is scalar
or a vector with dimension $D$).
This universality means that the details of the local forces
are unimportant; therefore the local interactions
can be simply modelled, through a schematic hamiltonian
supposed to hold the relevant symmetries
of the system.

Generally, the thermodynamic functions, such as the
free energy $F$ (from which the heat capacity 
$C = -T \partial^2F/\partial T^2$ 
can be derived) can be decomposed in a regular part,
and a singular part. The latter contain non-integer
exponents (such as some derivatives diverge) as a function
of $|T - T_c|$. Only the singular part with its
critical exponents are universal, but the
regular part could be dominant for some functions.

The 6 critical exponents, $\alpha$, $\beta$, $\gamma$, 
$\delta$, $\nu$ and $\eta$ are in fact not independent. From the
Widom \& Kadanoff scaling hypotheses, four relations can be
derived between them. These relations can be demonstrated
rigorously through renormalization theory. They are:
$$
2 \beta + \gamma = 2 - \alpha = 2 \beta \delta - \gamma
$$
$$
\gamma= \nu (2 - \eta)      
$$
$$
\nu d = 2 - \alpha
$$
where d is the space dimension of the problem.
There exists therefore only 2 independent critical exponents.

\bigskip
 \subsection{  The Ising Model}

The essential physical phenomena occuring when a system undergoes
a continuous phase transition are reproduced in simplified models,
that have played the role of prototypes. They consider a field
(order parameter) defined on a lattice of N sites, of dimension $d$.
For the Ising model of ferromagnets, the field is the spin value
at each site, i.e. $s_i$. We will consider only two possible
values for $s_i$, +1 or -1, so that the order parameter is a scalar
($D=1$).

The Ising model was resolved
for $d=1$ by Ising in 1925, for $d=2$ by Onsager in 1944, and with B non zero
only recently (Zamalodchikov 1989).There is not yet any 
analytical solution for $d=3$.

The Hamiltonian of the system can be written:
$$
H= 1/2 \sum_{ij} J_{ij} s_i s_j -B \sum_i s_i   
$$
where $B$ is the external magnetic field, and the interaction constant
$$
J_{ij} = J 
$$
if $i$ and $j$ are neighbouring sites, and 0 otherwise.
For ferromagnets, $J < 0$ and spins tend to align parallel to one another,
while $J > 0$ will correspond to anti-ferromagnets.

The partition function of the system in zero $B$ can be written as:
$$
Z = \sum_{si} exp(-\beta H) = \sum_{si} exp(-0.5\beta \sum_{ij} J_{ij} s_i s_j)
$$
Any physical quantity can then be derived, 
by summing over the possible configurations $\alpha$, 
such as the average energy:
$$
U = <E> = 1/Z \sum_\alpha E\alpha \, \, exp(-\beta E\alpha) = - \partial 
logZ/\partial \beta_V
$$
and the heat capacity in particular:
$$
<E^2> - <E>^2 = \partial^2logZ/\partial\beta^2_V = C_V /k\beta^2
$$
In the same way, correlation functions are expressed as
second derivatives of the partition function.
Since these summations are in general intractable analytically,
one has recourse to numerical techniques.

\bigskip
\subsection{Monte-Carlo Numerical Calculations}

The principle is to calculate directly the statistical
averages
$$
<X> = 1/Z \sum_{\alpha} X\alpha \, \, exp(-\beta E\alpha) 
$$
But the number of terms to compute grows exponentially
with the size of the system; let us consider for example the
Ising model in $d=2$, on a small lattice 3x3, the number
of configurations is
$2^9 = 512$, which induces reasonable computations. But
for a simple lattice 10x10, there are already
$2^{100}  \approx 1.3 \, 10^{30}$ configurations;
even with a sustained computational speed of 1 GFlop,
this will require more than 3 $10^{13}$ yr, i.e.
more than a thousand Hubble times.
Clearly, the method is to consider judicious samples
of the ensemble of configurations. There are configurations that 
are much less probable, and they will contribute negligibly.

The most simple method is the 
Metropolis one; it computes the statistical averages
in a certain number of steps (typically 10$^6$ steps or more),
each step corresponding to a configuration of the system.
Initially, the system is placed in a simple state (for instance
all spins aligned), but far from equilibrium. A certain number 
of steps are run without computing averages, to let the system
evolve far from the inital state, and the averages be independent
of it. The rule to change the configuration $\alpha$ 
to $\alpha'$ from one
step to the next is the simplest possible, i.e. 
reverse only one spin (so that the energy change between
the two states involves only the nearest neighbours), and
the probability that this state is effectively selected is:
$$
 P(\alpha \longrightarrow \alpha') =1
\; {\rm if}\; \Delta E = E\alpha'-E\alpha < 0$$
$$ {\rm and} \; 
P(\alpha \longrightarrow \alpha') = exp(-\beta \Delta E)
\; {\rm otherwise} \; $$
In this manner, the system is always in a very probable
state, with the probability corresponding exactly to
the actual one. The averages can therefore be taken
with equal weight for each of these states.

One big problem of this method, is that the
successive states are not independent, but correlated,
and the more so as we approach the critical point.
If the states were quasi-independent, the noise on
the computed averages will go down as $n_{step}^{-1/2}$,
but in fact it goes down slower.
The correlation time between configurations varies
according to the correlation length $\xi$ as
$$
\tau = \xi^z
$$
where $z$ is a power close to 2.
Near the critical point, where $\xi$ diverges,
there is a considerable slowing down of the method.

Other methods can cure this problem, such as the
Swendsen-Wang algorithm or the Wolff method.
The idea is to reverse blocks of spins simultaneously,
from one step to the next, and to reduce existing correlations,
in such a manner that existing clusters of aligned spins
are broken in smaller clusters, or different clusters.
 The algorithm consists of linking all aligned spins
by bonds, and so defining connected clusters. These
bonds are then destroyed randomly, but with probability
$exp(4 \beta J)$. The smaller clusters resulting from this
are re-oriented (up or down) randomly with equal probability.
In some cases, it can be shown that the power $z$ is then reduced
to zero.
In general computing the critical exponents through
direct Monte-Carlo methods is not precise enough, and 
renormalization techniques are required.

\bigskip
\subsection{ Renormalization }

Renormalization techniques were developped much earlier in
quantum field theory (e.g. Gell-Mann \& Low 1954), but their
application in statistical physics awaited the 1970s
(Wilson \& Kogut 1974; Wilson 1975, 1983), although 
Kadanoff (1966) presented already ideas anunciating 
how critical exponents could be extracted simply,
with great intuition about the physical processes
giving rise to critical phenomena.

\begin{figure}[t]
\psfig{figure=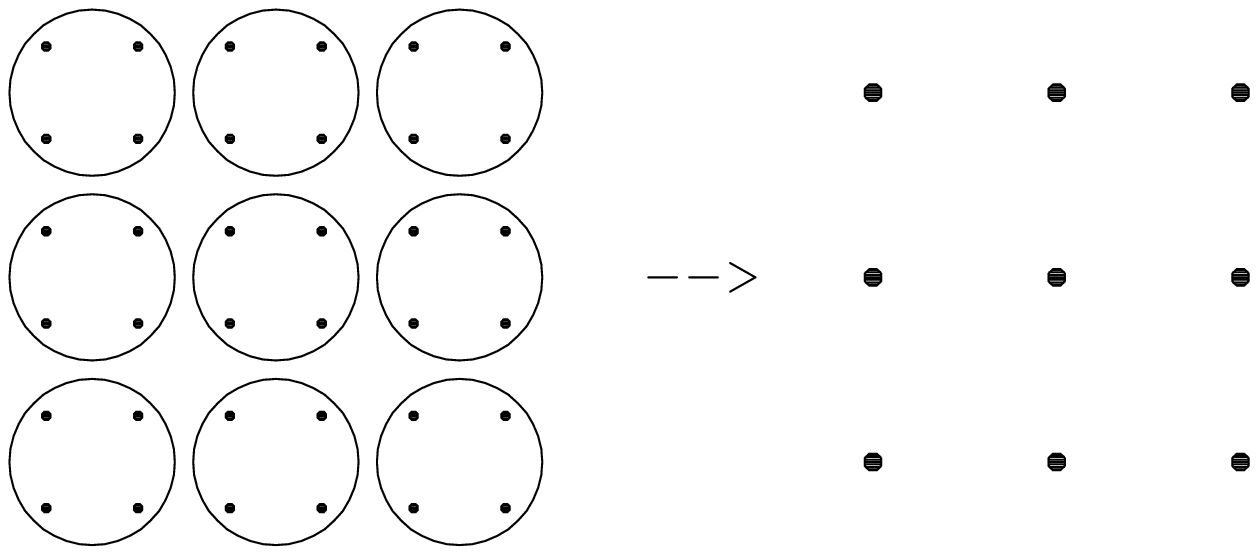,bbllx=5cm,bblly=55mm,bburx=20cm,bbury=12cm,width=10cm}
\caption{ Renormalization of a square lattice. Blocks of 4 sites are
replaced by one, dividing all scales by 2. }
\label{fig3}
\end{figure}

The principle of a renormalization in real space
(as opposed to conjugate space), is schematized in fig
\ref{fig3}. In the transformation of renormalization,
the scales are divided by a certain factor $k$, blocks
of a certain number of sites ($k^2$) are replaced by
one site, and since the system is scale-independent,
we should be able to find an hamiltonian for the blocks
which is of the same structure as the original one.
The new system is less critical than the previous one, since 
the correlation length $\xi$ has also been divided by $k$.
It is a way to reduce the number of degrees of freedom 
of the system.

If the transformation of renormalization is called {\bf R}
successive hamiltonians are related by 
$$
H_{n+1} = {\bf R} H_n
$$
At each step, a similar hamiltonian must be found, with however
different values of its parameters: coupling constant, temperature, etc..

The fixed points of the transformation, which obey
$$
H = {\bf R} H
$$
can be of various nature: attractive or repulsive (or mixed).
They are attractive when after several iterations, neighbouring
points are trapped there.
In many problems, there exist trivial points, in the asymptotic 
regimes of low ($T \longrightarrow  0$) 
or high ($T \longrightarrow  \infty$) temperatures.
But in between, there can exist fixed points, corresponding
to the critical points ($T = T_c$).

To give a very simple example, let us renormalize the Ising $d=1$
problem, which can be carried out easily (cf Lesne 1996).
The renormalization transformation will consist in considering
only the even-numbered sites, and drop the odd-numbered ones.
The scale is then divided by 2. The partition function can be written as:
$$
Z = \sum_j exp( K_0 \, s_{2j+1} (s_{2j} + s_{2j+2}))
$$
It is easy to demonstrate, since the spins can take only the values
+1 and -1, that:
$$
   exp( K_0 (s_3s_2 + s_3s_4)) = 2 exp (K_1) exp( K_1 s_2s_4)
$$
Then the partition function is exactly of the same form as the original
one, after the division by 2, with
$$
th(K_1) = th^2 (K_0) 
$$
(or $exp(2 K_1) = ch^2 (K_0)$).
Let us consider now the fixed points, where $K_1$ = $K_0$,
i.e. $th (K)$ = 0, or $th (K)$ = 1.
This arrives in the two extreme cases, $K \longrightarrow  0$ and 
$K \longrightarrow  \infty$.
$K$ is in fact the product of the coupling constant $J$, and $\beta$.
The first case corresponds, for a finite temperature, to zero coupling,
that is a purely thermal system, paramagnetic only; at zero $B$ there
is no correlation, and the system is not critical. Since the transform
of $K$ in this region is $\sim K^2$, and $\xi$ is divided by 2 at each
transform, we can deduce $\xi(K) = 2 \xi(K^2)$, and the behaviour
$\xi(K) \propto 1/|log(K)|$, i.e. $\xi \longrightarrow  0$.
The second fixed point corresponds to a zero temperature 
($K \longrightarrow \infty$).
Then we can show that $\xi(K) \propto exp(2 K)$, so $\xi$ diverges,
with large correlations. The system is truly critical.

In the general case,
it is quite complex to find the easiest way to build blocks,
and the new hamiltonian. A possible way is to Monte-Carlo
renormalize, i.e. 
represent the transformation matrix of {\bf R} as a statistical average
on the several possible choices.
The critical exponent is then the largest eigen value of this matrix. 

\bigskip
\subsection{ Mean Field Approximation}

This simple method to resolve the problem consists, for the Ising model of
ferromagnet as an example, to replace the action of the ensemble of
spins on a particular site, by the mean magnetic field at this position
(it is also independent of site by translational invariance). This
gives approximately when the system becomes critical (in general
$T_c$ is overestimated), and approximate values of the critical exponents
that do not depend on the spatial dimensionality (more exact when the
latter is large, or for infinite-range interactions).
For the 3D Ising model ($d=3$), the critical exponents can be
determined numerically to:
$\alpha$ = 0.107, $\gamma$ = 1.239, $\nu$ = 0.631 and $\eta$ = 0.037;
while in the mean field theory
$\alpha$ = 0, $\gamma$ = 1, $\nu$ = 0.5 and $\eta$ = 0; the approximation
appears therefore relatively bad.

\bigskip
\subsection{  Functional Integrals}

To efficiently use the renormalization methods, it is fruitful to 
go back to the field theory, domain in which the renormalization
was first introduced, and numerous techniques are available.
For example, in the case of the system of spins in a discrete lattice
$s_i$, the discrete configurations of the lattice are replaced by
a continuous field $s(\vec x)$. The summation on the sites $\Sigma_i$
are replaced by spatial integrals $\int d\vec x$; the hamiltonian
becomes a functional of the field $s(\vec x)$,
$$H(s,T) = \beta \int A(\vec s,\vec x) d\vec x$$
The summation over various configurations, to obtain for instance
the partition function $Z$, is replaced by a functional
integral, of variable the field  $s(\vec x)$:
$$ Z = \int_{\vec s} exp(-H(\vec s, T)$$
Often the calculations are more easy in the Fourier conjugate space,
the renormalised hamiltonian $H_1$ is defined of the same
form as the original one $H_0$, as
$$ exp(H_1) = \int_{\phi} exp(H_0(\Phi_0 + \phi))$$
The renormalised hamiltonian $H_1$ can be expressed as a perturbative
development, with the help of a diagrammatic analysis.

\bigskip
\section{  Statistical Self-Gravity }

As was shown is section 4, it appears that the self-gravitating
system is critical for a large range of the parameters, and
it is difficult to isolate a critical point, to identify diverging
behaviours. However, it is 
well known (Wilson 1975, Domb \& Green  1976), that
physical quantities diverge only for infinite volume systems,
at the critical point. Since our systems are also finite and
bounded, they only approach asymptotically the divergences.

\bigskip
\subsection{Identification of the Fractal Dimension}

If $ \Lambda $ measures the distance to the critical point,
 (in spin systems for instance, $ \Lambda $ is
proportional to $| T - T_c |$),
the correlation length  $ \xi $ diverges as, 
$$ 
\xi( \Lambda ) \sim  \Lambda^{-\nu} \; 
$$
and the specific heat (per unit volume) $ {\cal C} $ as,
$$
 {\cal C} \sim  \Lambda^{-\alpha}  \; 
$$
But in fact, for a finite volume system, all physical quantities are 
finite at the critical point.
When the typical size $R$ of the system 
is large, the  physical  magnitudes
take large values at the critical point, and
the infinite volume theory is used to treat finite size systems at
criticality. 
In particular, for our system, the correlation length provides the
relevant physical length $ \xi \sim R $, and we can write
$$
\Lambda \sim R^{-1/\nu} \; 
$$

Our system has the symmetries $d=3$ and $D=1$ (scalar field), which 
should indicate the universality class to which it corresponds.
It remains to identify the correponding operators. Already in the previous
sections, it was suggested that the field $\phi$ corresponds
to the potential, and the mass density  
$$
m\, \rho({\vec x}) = m\, \,  e^{\phi({\vec x})}
$$
can be identified with the  energy density in the renormalization
group (also called the `thermal perturbation operator').

We note that the state of zero density (or zero fugacity),
corresponds to a singular point, around which we develop the
physical functions (and we choose $\Lambda$ accordingly).
At this point $ \mu^2/T_{eff} = 0 $, the partition function $ {\cal Z}
$ is singular
$$
  \Lambda \equiv   {{\mu^2}\over{T_{eff}}} = z\,
  \left({{2\pi mkT}\over{h^2}}\right)^{3/2} \; 
$$
i.e., the critical point $  \Lambda = 0 $ corresponds to zero
fugacity $z$. We can write $ {\cal Z} $ as a function of the 
action $S^*$ at the critical point
$$
{\cal Z}(\Lambda) = 
 \int\int\;  {\cal D}\phi\;  e^{ -S^* + \Lambda
\int d^3x  \; e^{\phi({\vec x})}\;}
$$

Let us now decompose log$ {\cal Z} $  in its singular and regular parts
(we know that the second derivative will give the heat capacity
that diverges in $\Lambda^{-\alpha}$)

$$
{1 \over V} \; \log{\cal Z}(\Lambda) = {K \over{(2-\alpha)(1-\alpha)}}\;
\Lambda^{2-\alpha} + F(\Lambda) \; 
$$

where $  F(\Lambda) $ is an analytic
function of $ \Lambda $ around the origin 
$$ 
F(\Lambda) = a \; \Lambda + \frac12 \, b  \; \Lambda^2 + \ldots 
$$ 
 $ V = R^3 $ stands for the volume and $ K, \; a $ and $ b $ are constants.

By derivation with respect to $ \Lambda $ 
$$
{1 \over V} \;{{\partial}\over{\partial\Lambda}}\log{\cal Z}(\Lambda)=
a +  {K \over{1-\alpha}}\,
\Lambda^{1-\alpha}
+ \ldots = {1 \over V} \int d^3x  \; <e^{\phi({\vec x})}>\; 
$$
Now, using the standard relation $ \alpha = 2 - \nu d = 2 - 3 \nu$,
this gives
$$
{{\partial}\over{\partial\Lambda}}\log{\cal Z}(\Lambda)= V \, a +
 {K \over{1-\alpha}}\, R^{1/\nu} + \ldots
$$
The mass contained in a region
of size $ R $ is
$$
M(R) = m  \int^R e^{\phi({\vec x})} \; d^3x  
$$
$$
<M(R)> =  m  \, V \, a +m  \, {K \over{1-\alpha}}\; R^{ \frac1{\nu}} +
\ldots
$$
We already see that, apart a possible constant density, the 
average mass obeys a singular power-law.

The 2-points density correlator varies as
$$
C({\vec r_1},{\vec r_2})\sim |{\vec r_1} -{\vec r_2}|^{\frac2{\nu} -6} 
$$
The perturbative calculation  matches
with this result for $ \nu = \frac12 $, the mean field
value for the exponent $ \nu $. 
Pursuing further to the 
second derivative of $ \log{\cal Z}(\Lambda) $
with respect to $\Lambda$ 
$$
{{\partial^2}\over{\partial\Lambda^2}}\log{\cal Z}(\Lambda)= V\left[
\Lambda^{-\alpha} \, K + b  + \ldots  \right] 
$$
$$
{{\partial^2}\over{\partial\Lambda^2}}\log{\cal Z}(\Lambda)=
\int d^3x\; d^3y\; C({\vec x},{\vec y}) \sim R^D \int^R 
{{ d^3x}\over{x^{6 - 2d_H}}}  \sim \Lambda^{-2}\sim R^D \;
\Lambda^{-\alpha} 
$$
we can find the mass fluctuations and corresponding dispersion:
$$
(\Delta M(R))^2 \equiv  \; <M^2> -<M>^2 \sim
\int d^3x\; d^3y\; C({\vec x},{\vec y}) \sim R^{2/\nu}
$$
$$
\Delta M(R)  \sim  R^{1/\nu}
$$
This is the definition relation of the fractal, with dimension $d_H$,
and the scaling exponent $\nu$ can be identified with the inverse
Haussdorf dimension of the system
$$
d_H = \frac1{\nu} 
$$
The velocity dispersion follows
$$
\Delta v \sim R^{q} 
$$
with 
$$
q =\frac12\left(\frac1{\nu} -1\right) =\frac12(d_H -1)  \;
$$

\bigskip
\subsection{Numerical Values}

The scaling exponents $ \nu , \; \alpha $ can
be computed through the renormalization group approach. The case of a 
single component (scalar) field has been extensively studied 
in the literature (Hasenfratz \& Hasenfratz 1986, Morris 1994a,b). 
Very probably, there is a
unique, infrared stable fixed point in 3D: the
Ising  model fixed point. Such  non-perturbative fixed point is
reached in the long scale regime independently of the initial shape of
the interaction ($\phi$).
For the Ising model $d=3$, the exponents are:
$\nu$ = 0.631, from which we deduce $d_H$ = 1.585,  $\eta$ =
0.037, and $\alpha$= 0.107.

 The value of the dimensionless coupling constant $g^2 = \mu T_{eff}$
should decide whether the fixed point chosen by the system is the
mean field (weak coupling) or the Ising one (strong coupling).
At the tree level, we
estimate  $g \approx \frac{5}{\sqrt{N}}$, where
$N$ is the number of points in a Jeans volume $d_J^3$. The coupling
constant appears then of the order of 1,
and we cannot settle this question
without effective computations of the renormalization group equations.
In any case,  
both the Ising  and mean field values are
in  agreement with the astronomical
observations (the mean field exponents are 
$\nu$ = 0.5 $d_H$ = 2, $\eta$ =0 and $\alpha$ = 0).

\bigskip
\subsection{Important Differences with the Spin Models }

For the gravitational gas we find scaling behaviour for a full range of
temperatures and couplings, while
for spin models scaling only appears 
at the critical value of the temperature.
At $T = T_c$, the correlation length $\xi$ is infinite, and the 
theory is massless.
In fact, since the spin systems are not infinite,
even in the critical
domain,  $ \xi $ is finite and the correlation functions
decrease as $ \sim e^{ - r/\xi} $ for large distances $ r $
(only $\xi$ is large, of the order of the system size).
Fluctuations of the relevant operators support perturbations which can
be interpreted as massive excitations. 

Such (massive) behaviour does not appear for the gravitational gas. The density
correlators scale, exhibiting power-law behaviour. This feature is
connected with the scale invariant character of the Newtonian force
and its infinite range.

\bigskip
\subsection{Observational Tests}

It was shown that the predicted fractal dimension of a self-gravitating
critical medium is compatible with that of the astrophysical
applications (interstellar medium, galaxies), within the observational
uncertainties. However, there are other predictions of the theory, that
could be checked. 
 It is well known in critical phenomena that there
exists two independent critical exponents: the second one
concerns the correlations of the potential, corresponding to the field $\phi$
$$
<\phi({\vec r})\phi(0)> \sim r^{-1-\eta} \; 
$$

The potential is then predicted to vary as a power law with size,
with a slope $-1/2 -\eta/2$ = -0.518.
It is not easy to observe directly this quantity, but tracers of 
the potential could be obtained through light rays that are
deviated by astrophysical masses. As for the ISM, for example,
the potential fluctuations could be traced by the micro-lensing
produced on the light of remote quasars, while they are crossing the ISM
of intervening galaxies, at high redshift. Time fluctuations of the
quasars flux is regularly observed, and interpretation in terms
of micro-lensing has been proposed (Lewis et al 1993).

\bigskip
\section{ Conclusion}

We have emphasized the existence of two astrophysical fractals,
the interstellar medium, with structures ranging from 10 AU to
100 pc, and the large-scale structures of galaxies, from 50 kpc
to 150 Mpc at least. The first one is in statistical equilibrium,
while the second one is still growing to larger scales. In both
cases, we can describe these media as developping large-scale
fluctuations with large correlations as is familiar in critical
phenomena.
We propose that in both cases, self-gravity is the main force
governing these fractal structures.

The statistical thermodynamic approach is developped, and
it is shown that the phenomenological potential, which is
in $1/r$ between two cutoffs (at small and large-scale),
can be described by a scalar field theory.
 We use the renormalization group methods 
for this scale-independent system, to find
the universality class of the problem. The fractal
dimension, and the potential correlations exponent
can be derived from the critical
exponents of the Ising $d=3$ model.
The stability of the results with respect to perturbations has
been studied; the results are quite robust with respect to
perturbations of external forces.
Of course,  if the external energy is large with respect to the 
self-gravitational
energy of the gas (in the vicinity of violent star-formation in the ISM
for instance), then the hierachical structure will be destroyed.
The observed fractal dimensions are compatible with the predictions,
and other observational tests are proposed.

\end{document}